\documentclass[letterpaper,conference]{IEEEtran}
\usepackage{cite}
\usepackage{amsmath,amssymb,amsfonts}
\usepackage{algorithmic}
\usepackage{graphicx}
\usepackage{textcomp}
\usepackage[table]{xcolor}
\usepackage[caption=false]{subfig}
\usepackage{amsmath}
\usepackage{multirow}
\usepackage{soul}
\usepackage[]{nohyperref}  
\usepackage{url}  

\def\BibTeX{{\rm B\kern-.05em{\sc i\kern-.025em b}\kern-.08em
	T\kern-.1667em\lower.7ex\hbox{E}\kern-.125emX}}
\begin{document}

\title{Smart Space Environments: Key Challenges and Innovative Solutions}
	\author{\IEEEauthorblockN{Ramakant Kumar}
	}

\maketitle

\begin{abstract}
The integration of LoRaWAN (Long Range Wide Area Network) technology with both active and passive sensors presents a transformative opportunity for the development of smart home systems. This paper explores how active sensors, such as motion detectors and ultrasonic sensors, and passive sensors, including temperature and humidity sensors, work together to enhance connectivity and efficiency within diverse environments while addressing the challenges of modern living. By leveraging LoRaWAN’s long-range capabilities and low power consumption, the proposed framework enables effective data transmission from remote sensors, facilitating applications such as smart agriculture, environmental monitoring, and comprehensive home automation. Active sensors emit energy to detect changes in their surroundings, providing real-time data crucial for security and automation, while passive sensors capture ambient energy to monitor environmental conditions, ensuring resource efficiency and user comfort. The synergy between LoRaWAN and these various sensor types promotes innovation, contributing to a more responsive and sustainable living experience. Furthermore, this research highlights the adaptability of the proposed system, allowing for seamless integration of new devices and advanced functionalities. As the landscape of smart home technology continues to evolve, ongoing research in this area will yield advanced solutions tailored to user needs, ultimately paving the way for smarter, safer, and more efficient living environments.
\end{abstract}

\begin{IEEEkeywords}
LoRaWAN, smart campuses, communication protocols, network optimization, spreading factor, transmission power, Edge computing, Fog computing, Cloud computing, IoT, mobility management, signal propagation, path loss, smart space solutions.
\end{IEEEkeywords}

\section{Introduction}
Advancements in low-cost sensors and state-of-the-art hardware have catalyzed the development of smart campuses, where diverse devices and systems interconnect to optimize operational efficiency and enhance user experience. These campuses integrate a multitude of Internet of Things (IoT) devices, including sensors, actuators, and controllers, which work collaboratively to enable seamless communication and automation across various applications, such as energy management, security, and resource monitoring~\cite{7123563, conf/wcnc/0001G24, journals/tpds/MishraGBD24}.

In this context, both wired and wireless communication technologies are pivotal, providing a reliable foundation for data transmission across devices. Wireless communication, in particular, has gained significant traction due to its flexibility and scalability. Technologies span from short-range protocols like Wi-Fi and Bluetooth, used for high-bandwidth applications within confined areas, to long-range protocols such as LoRa and LoRaWAN, which support low-power, wide-area communication. These long-range solutions enable smart campus systems to monitor and manage resources across large geographical areas with minimal infrastructure requirements, thereby maximizing energy efficiency and minimizing operational costs. LoRa, a prominent long-range wireless communication technology, is especially valuable for supporting real-time monitoring and control in campus-wide IoT deployments, facilitating a robust ecosystem that drives the vision of intelligent, connected spaces.

This paper begins by introducing the fundamental concepts essential to the design and deployment of smart space environments, starting with an overview of sensor types, including both active and passive sensors. Active sensors, which emit energy to detect environmental changes, and passive sensors, which rely on ambient energy for detection, each play a vital role in gathering data from diverse campus settings. These sensors serve as the backbone for data acquisition, enabling the continuous monitoring of environmental and operational parameters.

We then discuss the various communication protocols, distinguishing between wired and wireless methods, as well as short-range and long-range protocols. Short-range protocols, such as Bluetooth and Wi-Fi, support high-speed data transfer over confined distances and are particularly suitable for indoor applications that require fast and reliable connectivity. Conversely, long-range protocols like LoRa and LoRaWAN facilitate low-power, wide-area communication, making them ideal for large-scale outdoor applications across campus environments. By utilizing these communication technologies, smart spaces can achieve seamless connectivity and data flow across interconnected devices, thereby enhancing real-time monitoring, control, and automation capabilities~\cite{9164991,5558084,6054047}.

Following this foundational discussion, we delve into smart space solutions that leverage advanced computing paradigms such as Edge, Fog, and Cloud computing. Edge computing, by processing data closer to the source, reduces latency and minimizes bandwidth usage, thereby enabling faster response times for critical applications~\cite{7488250}. Fog computing extends these capabilities by providing a distributed computing layer between the Edge and Cloud, thus offering scalable processing power and storage closer to campus devices. Finally, Cloud computing delivers extensive computational resources and centralized data storage, supporting complex data analytics and machine learning tasks that enhance campus-wide operational efficiency. Together, these computing paradigms create a cohesive infrastructure that empowers smart spaces to operate intelligently and autonomously, effectively managing data-intensive tasks while meeting the dynamic needs of modern campus environments~\cite{7123563,8016573,7498684}.

\section{Sensors}
Sensors are fundamental components in smart space environments, enabling real-time data acquisition from physical surroundings. These devices convert physical parameters—such as temperature, humidity, motion, and light—into electrical signals, which are then processed and analyzed to support decision-making and automation. Sensors can be categorized broadly into active and passive types. Active sensors emit energy to detect and measure environmental changes, often being used in applications like security monitoring and object detection. Passive sensors, on the other hand, do not emit energy but instead rely on naturally occurring energy sources, making them ideal for environmental monitoring due to their lower power consumption~\cite{journals/csur/MishraG23,journals/wpc/ChopadeGD23,conf/iccps/0012G023}.

In addition to sensors, communication protocols play a crucial role in facilitating data transmission across connected devices within smart spaces. Communication protocols can be classified as wired or wireless, and further into short-range and long-range categories. Short-range protocols, such as Bluetooth and Wi-Fi, are widely used for high-speed data transmission over short distances and are suitable for applications within confined areas. Long-range protocols like LoRa and LoRaWAN, known for their energy efficiency and extended coverage, support low-power communication over vast areas and are particularly advantageous for outdoor and large-scale campus applications~\cite{7123563,9130098,conf/infocom/MishraGD22}.

Together, sensors and communication protocols form the backbone of smart spaces, enabling seamless connectivity and data sharing across diverse devices and systems. This interconnected network supports a wide range of applications, from energy management and security to environmental monitoring, fostering intelligent, responsive, and efficient campus environments.

\subsection{Active Sensors in Smart Home Systems}
Active sensors play a pivotal role in smart home systems by actively emitting energy, such as electromagnetic or acoustic waves, toward a target to detect and measure environmental changes. By sending out signals and then analyzing the returned data, these sensors can measure attributes like distance, motion, and even object size. In smart home applications, active sensors enhance automation and security, enabling the home to respond dynamically to changes within the environment.

\subsubsection{Working Principle}
Active sensors operate by emitting a signal that interacts with objects or surfaces in the environment. When the emitted signal encounters an object, part of the signal reflects back to the sensor. The sensor then measures specific characteristics of the reflected signal—such as the time taken for the return, the change in frequency, or the signal strength—to interpret information about the object’s distance, movement, or other properties. This method allows for continuous, real-time monitoring and data acquisition, making active sensors ideal for responsive smart systems.

\subsubsection{Example Applications in a Smart Home}
\begin{itemize}
    \item \textbf{Radar Motion Sensors for Security}
    \begin{itemize}
        \item \textit{Operation}: Radar sensors emit microwave or radio waves, which bounce off any object in their path. When an intruder enters the home, the radar sensor detects movement by analyzing the changes in the reflected signals.
        \item \textit{Application}: Used for detecting unauthorized entry, radar sensors enable security alerts and can even trigger recording in security cameras. They can also differentiate between objects moving at different speeds, such as distinguishing between a pet and a human intruder.
    \end{itemize}
    
    \item \textbf{LiDAR for Intruder Detection and Room Mapping}
    \begin{itemize}
        \item \textit{Operation}: LiDAR sensors emit pulses of laser light and measure the time it takes for each pulse to return. By calculating the time delay, the sensor builds a 3D map of the surroundings.
        \item \textit{Application}: In a smart home, LiDAR can assist with security and spatial awareness, mapping the layout of rooms and detecting unexpected intrusions. It can also be used to improve the navigation of autonomous vacuum cleaners by enabling them to recognize obstacles and efficiently map rooms.
    \end{itemize}
    
    \item \textbf{Ultrasonic Sensors for Smart Lighting Control}
    \begin{itemize}
        \item \textit{Operation}: Ultrasonic sensors emit high-frequency sound waves and measure the time taken for the echoes to return. By analyzing the frequency and time delay, the sensor can detect motion and determine the distance to an object.
        \item \textit{Application}: Ultrasonic sensors can be used to automate lighting by detecting when someone enters or leaves a room. For instance, when movement is detected, the lights automatically switch on, and they switch off after a certain period of inactivity, optimizing energy consumption.
    \end{itemize}
\end{itemize}

In each scenario, active sensors provide a responsive, real-time monitoring capability that enhances both the efficiency and security of smart home systems. Their ability to interact dynamically with their surroundings allows them to detect subtle changes and respond instantly, creating a smarter and safer living environment.

\subsection{Passive Sensors}
Passive sensors, unlike active sensors, do not emit energy but instead detect natural energy from the environment. They are often used to monitor environmental conditions by capturing information such as heat, light, or sound emitted by objects or surroundings. Examples of passive sensors include temperature sensors, photodetectors, and infrared sensors, which detect variations in ambient energy to provide data on factors like temperature, humidity, and light levels. These sensors are crucial for applications focused on resource efficiency, environmental monitoring, and user comfort, as they consume less power and offer longevity, making them well-suited for large-scale deployment in smart campuses.

\section{Wireless Networking Protocol}
Wireless networking protocols play a crucial role in enabling communication between devices across various applications, including the Internet of Things (IoT). These protocols can be broadly classified into two main categories: short-range and long-range communication protocols, depending on their operational range, power requirements, and data rate capabilities. This section provides an overview of these categories and highlights several widely used protocols within each type.

\subsection{Passive Sensors in Smart Home Systems}
Passive sensors, unlike active sensors, do not emit energy but instead detect natural energy from the environment. They are often used to monitor environmental conditions by capturing information such as heat, light, or sound emitted by objects or surroundings. Examples of passive sensors include temperature sensors, photodetectors, and infrared sensors, which detect variations in ambient energy to provide data on factors like temperature, humidity, and light levels. These sensors are crucial for applications focused on resource efficiency, environmental monitoring, and user comfort, as they consume less power and offer longevity, making them well-suited for large-scale deployment in smart homes.

\subsubsection{Working Principle}
Passive sensors operate by detecting and measuring the natural energy emitted or reflected by objects and the surrounding environment, without emitting any signals of their own. These sensors capture variations in ambient energy—such as thermal energy from heat sources, light from natural or artificial sources, or sound from nearby activity—and convert this data into electrical signals. This approach enables passive sensors to function continuously with minimal power consumption, making them ideal for applications that require constant monitoring over long periods.

\subsubsection{Example Applications in a Smart Home}
\begin{itemize}
    \item \textbf{Temperature Sensors for Climate Control}
    \begin{itemize}
        \item \textit{Operation}: Temperature sensors measure the ambient thermal energy in a room and convert it into an electrical signal. This data can be used to determine the current temperature and adjust it according to user preferences.
        \item \textit{Application}: In a smart home, temperature sensors are essential for optimizing heating and cooling systems. The smart thermostat can adjust the HVAC system to maintain comfortable indoor temperatures, reducing energy consumption and costs by ensuring the system only operates when necessary.
    \end{itemize}
    
    \item \textbf{Photodetectors for Smart Lighting}
    \begin{itemize}
        \item \textit{Operation}: Photodetectors sense the intensity of light in an area by detecting natural or artificial light levels. They generate an electrical signal proportional to the light intensity.
        \item \textit{Application}: In a smart home, photodetectors allow lighting systems to adjust automatically based on the ambient light. For instance, they can dim or brighten lights depending on the time of day, reducing energy consumption by ensuring lights are used only when needed.
    \end{itemize}
    
    \item \textbf{Infrared Sensors for Motion Detection}
    \begin{itemize}
        \item \textit{Operation}: Infrared sensors detect the heat emitted by objects, such as people, by measuring infrared radiation. They can identify changes in infrared energy to detect motion.
        \item \textit{Application}: Commonly used for security and automation, infrared sensors can detect when someone enters or leaves a room, triggering events such as turning on lights or activating security cameras. This ensures user comfort and security without the need for high-power sensors.
    \end{itemize}
\end{itemize}

In each of these smart home applications, passive sensors contribute to a more energy-efficient and user-friendly environment. Their low power consumption and longevity make them an excellent choice for applications requiring ongoing monitoring and sustainability in a smart home system.

\subsection{Active and passive sensors with Communication Protocols}

\subsection{Examples of Sensors with Communication Protocols in Smart Home Applications}

\noindent Here are some examples of how different sensors operate with various communication protocols in a smart home environment:

\subsubsection{Temperature Sensor with Wi-Fi Communication}
\begin{itemize}
    \item \textbf{Working Principle:} Temperature sensors detect ambient thermal energy and convert it into an electrical signal that represents the temperature level. These sensors are passive, using minimal energy to monitor room conditions.
    \item \textbf{Communication:} Wi-Fi, with its higher data rate and longer range (up to 100 meters indoors), enables temperature sensors to transmit data to a central hub or directly to a smartphone app.
    \item \textbf{Application:} In a smart home, temperature sensors with Wi-Fi are often used in smart thermostats to monitor and control heating and cooling systems. The data can be accessed in real-time through an app, allowing homeowners to adjust the temperature remotely for comfort and energy savings.
\end{itemize}

\subsubsection{Light Sensor (Photodetector) with Bluetooth Communication}
\begin{itemize}
    \item \textbf{Working Principle:} Light sensors, or photodetectors, measure the intensity of ambient light by detecting photons and converting this light into an electrical signal proportional to the brightness level.
    \item \textbf{Communication:} Bluetooth, with a range of up to 100 meters and low energy usage, allows light sensors to send data to nearby devices like smartphones, smart bulbs, or central hubs.
    \item \textbf{Application:} In smart lighting systems, light sensors with Bluetooth can adjust indoor lighting based on natural light levels. For instance, they can dim lights when sufficient natural light is present or brighten them during dim conditions, improving energy efficiency without the need for manual adjustments.
\end{itemize}

\subsubsection{Gas Sensor with LoRa Communication}
\begin{itemize}
    \item \textbf{Working Principle:} Gas sensors detect specific gases (such as carbon dioxide or methane) by measuring changes in resistance when the gas interacts with the sensor material, producing an electrical signal that indicates gas concentration.
    \item \textbf{Communication:} LoRa (Long Range) provides low-power, long-range communication (up to several kilometers in open areas), making it ideal for gas sensors that need to transmit data across distances without frequent recharging.
    \item \textbf{Application:} In smart homes, gas sensors with LoRa communication are used to monitor air quality, especially in areas like kitchens. These sensors can alert users remotely via smartphone apps in case of dangerous gas levels, enhancing safety with real-time monitoring.
\end{itemize}

\subsubsection{Sound Sensor (Microphone) with NFC Communication}
\begin{itemize}
    \item \textbf{Working Principle:} Sound sensors, such as microphones, capture audio waves in the environment and convert them into electrical signals. These sensors are passive and detect natural sounds without emitting energy.
    \item \textbf{Communication:} NFC (Near Field Communication), with a very short range (up to 10 cm), can transmit data when in close proximity to another NFC-enabled device~\cite{conf/wowmom/MishraKGSDSP21,7803607,conf/infocom/KumariGD21,7377400,8326735}.
    \item \textbf{Application:} In a smart home, a sound sensor with NFC might be used for specific, short-range applications like activating devices with a voice command or controlling access to a room. For example, a door lock could be controlled by an NFC-enabled sound sensor that detects a specific sound pattern or voice command.
\end{itemize}

\subsubsection{Vibration Sensor (Accelerometer) with Z-Wave Communication}
\begin{itemize}
    \item \textbf{Working Principle:} Vibration sensors measure changes in acceleration along different axes, detecting movement or vibrations. They use a piezoelectric material or MEMS (Micro-Electro-Mechanical Systems) to detect shifts in position or orientation.
    \item \textbf{Communication:} Z-Wave, a low-power protocol with a range of up to 100 meters, is well-suited for smart home devices and can connect multiple nodes reliably.
    \item \textbf{Application:} In a smart home, vibration sensors with Z-Wave communication can be used on doors or windows to detect unauthorized entry, triggering alerts or connecting to security systems. This configuration is low-power and can remain active for extended periods, enhancing home security~\cite{Wang2021,Charef2023,Samariya2023,Sikder2023,Rodriguez2023}.
\end{itemize}

\subsubsection{Infrared Sensor with Infrared Communication}
\begin{itemize}
    \item \textbf{Working Principle:} Infrared (IR) sensors detect infrared radiation emitted by nearby objects. When a warm body, such as a human, enters the detection range, the IR sensor detects the change in radiation.
    \item \textbf{Communication:} Infrared communication transmits data using infrared light up to 5 meters, typically through line-of-sight connections between devices~\cite{Kea2023,Shubyn2023,Qin2020,Mohammadi2023}.
    \item \textbf{Application:} IR sensors and communication are common in smart remotes, where users can control appliances by aiming the remote at a device. In a smart home, IR sensors may also be used for motion detection in lights or for controlling IR-compatible home devices like TVs or air conditioners.
\end{itemize}

\noindent Each of these combinations of sensors and communication protocols provides specific functionalities that enhance security, comfort, and energy efficiency in smart home environments.

\begin{table*}[h!]
    \centering
    \caption{Sensors and Communication Protocols for Smart Home Applications}
    \begin{tabular}{|c|c|c|c|c|}
        \hline
        \textbf{Sensor Type} & \textbf{Communication Protocol} & \textbf{Range} & \textbf{Data Rate} & \textbf{Reference} \\
        \hline
        Temperature Sensor & Wi-Fi & 50-100 meters indoors & Up to 1 Gbps & \cite{conf/wcnc/ThangadoraiSGK24} \\
        \hline
        Light Sensor (Photodetector) & Bluetooth & Up to 100 meters & Up to 3 Mbps & -- \\
        \hline
        Gas Sensor & LoRa & Several kilometers & Low & -- \\
        \hline
        Sound Sensor (Microphone) & NFC & Up to 10 cm & Very Low & \cite{9311219} \\
        \hline
        Vibration Sensor (Accelerometer) & Z-Wave & Up to 100 meters & 9.6 to 100 kbps & \cite{7745306} \\
        \hline
        Infrared Sensor & Infrared (IR) & Up to 5 meters & Up to 4 Mbps & -- \\
        \hline
        Proximity Sensor & RFID & Up to 100 meters (active tags) & Very Low & \cite{1589116,7123563} \\
        \hline
        Motion Sensor & Zigbee & 10-100 meters & 250 kbps & \cite{4460126} \\
        \hline
        Environmental Sensor (e.g., Temperature, Humidity) & M-Bus & Up to 1 km & Variable & -- \\
        \hline
    \end{tabular}
    \label{tab:sensors_protocols}
\end{table*}

\begin{table*}[h!]
    \centering
    \caption{Sensors with Communication for IoT Applications}
    \begin{tabular}{|p{3cm}|p{3cm}|p{3cm}|p{3cm}|p{3cm}|p{1cm}|}
        \hline
        \textbf{Protocol} & \textbf{Range} & \textbf{Data Rate} & \textbf{Suitable Applications} & \textbf{Associated Sensors} & \textbf{Reference} \\
        \hline
        Sigfox & Up to 50 km in rural areas & ~100 bps & IoT devices transmitting small data infrequently & Environmental sensors, Smart meters & \cite{8480255} \\
        \hline
        NB-IoT (Narrowband IoT) & Up to 35 km & Moderate & Smart cities, utility metering & Water quality sensors, Gas sensors & \cite{8170296,8502812} \\
        \hline
        LTE-M (Long Term Evolution for Machines) & Similar to NB-IoT & Up to 1 Mbps & Connected vehicles, wearables & GPS sensors, Health monitoring devices & \cite{7880946} \\
        \hline
        4G (Fourth Generation) & Extensive & Up to 1 Gbps & Mobile broadband, IoT applications & Surveillance cameras, Smart home devices & -- \\
        \hline
        5G (Fifth Generation) & Extensive & Up to 10 Gbps & Critical IoT, high-data applications & Autonomous vehicle sensors, High-definition cameras & -- \\
        \hline
        Mioty & Up to 10 km & Low & Industrial IoT, smart city applications & Industrial sensors, Smart meters & \cite{10176016} \\
        \hline
        Satellite & Global coverage & Variable & Remote monitoring (agriculture, forestry) & Weather sensors, Environmental monitoring devices & -- \\
        \hline
    \end{tabular}
    \label{tab:long_range}
\end{table*}

\subsection{Challenges in Communication with Sensors for Smart Home}
Short-range communication protocols are widely used for various applications, particularly in personal area networks and smart home devices. The main challenge for these protocols is their limited range, which typically spans from a few centimeters to about a hundred meters, depending on the protocol~\cite{5567086,9130098,7460727,9479778,1321026,9164991}. This limited distance can be a significant constraint in larger environments, such as warehouses or outdoor settings, where multiple devices need to communicate over extended areas. Physical obstacles like walls and furniture can degrade signal strength, leading to interruptions in communication.

Despite these limitations, short-range protocols are designed to be energy-efficient, allowing devices to operate effectively for extended periods without frequent battery replacements or recharging. This is essential for applications in smart homes that utilize numerous sensors and devices.

Active sensors, such as radar and LiDAR, emit energy towards a target to detect changes in the environment. They work by sending out signals and analyzing the returned data, which allows for accurate measurements of distance and movement. In smart homes, active sensors can be used for applications like movement detection and security surveillance.

Passive sensors, on the other hand, do not emit energy but instead detect natural energy emitted by objects in the environment. They capture information such as temperature, light, or sound. For instance, a temperature sensor measures ambient temperature variations, which can be critical for HVAC systems in smart homes. Passive sensors are often more energy-efficient, making them suitable for long-term monitoring without frequent maintenance.

Long-range communication protocols, such as Sigfox and NB-IoT, are designed to facilitate connectivity over extended distances but often face challenges related to energy consumption and infrastructure requirements. Nevertheless, these protocols are crucial for applications like smart city initiatives and remote asset tracking, where connectivity over vast areas is essential.

\section{LoRaWAN for Smart Home with Sensors}

LoRaWAN (Long Range Wide Area Network) is emerging as a transformative technology for smart home applications, enabling efficient communication between various sensors deployed in residential environments. Its ability to connect devices over long distances while maintaining low power consumption makes it particularly well-suited for smart homes, where numerous sensors and devices need to communicate seamlessly.

In the context of smart homes, active sensors play a crucial role in monitoring and controlling various aspects of the environment. These sensors, such as motion detectors and LiDAR, actively emit signals to measure parameters like distance and movement. By integrating active sensors with LoRaWAN, homeowners can achieve real-time data collection and analysis, enhancing security systems and automating lighting based on occupancy.

Conversely, passive sensors are equally important in a smart home ecosystem. These sensors, which include temperature sensors, humidity sensors, and light detectors, do not emit signals but instead detect natural energy from their surroundings. For instance, temperature sensors can monitor the ambient temperature to optimize heating and cooling systems, while light sensors can adjust indoor lighting based on the natural light available. When connected to LoRaWAN, passive sensors provide valuable data that can be processed to improve energy efficiency and user comfort in the home.

The combination of LoRaWAN with both active and passive sensors enables a comprehensive smart home system that not only enhances convenience and security but also promotes sustainability. Active sensors, such as motion detectors, sound sensors, and infrared cameras, provide immediate feedback and real-time data, allowing homeowners to respond promptly to security threats or environmental changes. For instance, a motion sensor can trigger lighting or alarms when movement is detected, enhancing security while conserving energy by only activating systems when needed.

Passive sensors, including temperature, humidity, and light sensors, play a vital role in optimizing energy consumption within smart homes. By continuously monitoring environmental conditions, these sensors can adjust heating, ventilation, and air conditioning (HVAC) systems to maintain comfort while minimizing energy use. For example, smart thermostats equipped with passive temperature sensors can learn user preferences over time, leading to automated adjustments that save energy without sacrificing comfort. LoRaWAN’s long-range capabilities ensure that even remote sensors can transmit data effectively, facilitating smart agriculture practices and environmental monitoring, where sensor deployment may be challenging due to distance and power constraints~\cite{conf/wcnc/PandeyKG024,conf/icc/KumariGDB23}. This is particularly beneficial for monitoring soil moisture levels, temperature, and air quality in agricultural settings, where traditional networking solutions may fall short. Farmers can receive real-time data on crop conditions, allowing for timely interventions that optimize yields and reduce resource waste. Additionally, its low power consumption allows devices to operate for extended periods on small batteries, making it ideal for long-term installations in hard-to-reach areas~\cite{conf/wowmom/KumariGDD22,7815384,conf/mswim/KumariGM021}. This advantage is crucial for both smart home applications and large-scale environmental monitoring projects, where replacing batteries can be labor-intensive and costly. LoRaWAN’s energy-efficient architecture enables devices to last for years without maintenance, thus ensuring uninterrupted data collection. Furthermore, the scalability of LoRaWAN networks allows for easy integration of new devices as smart home ecosystems grow. Homeowners can start with a few sensors and gradually expand their system by adding more active and passive sensors tailored to their evolving needs. This flexibility fosters innovation and encourages the adoption of smart technologies, enhancing the overall user experience.

Integrating LoRaWAN with various sensors in smart homes creates a robust framework that addresses the demands of modern living while ensuring efficiency and adaptability in diverse environments~\cite{conf/icc/KumariGD20,journals/cem/GhoshMGSR22}. This integration enables seamless communication between devices, allowing homeowners to monitor and control their environments more effectively. For example, smart temperature and humidity sensors can work in conjunction with HVAC systems, automatically adjusting settings based on real-time data to optimize energy use and maintain comfort.

The synergy of technology fosters innovation in home automation, contributing to a smarter, more responsive living experience~\cite{6780609,9166711,5558084,6054047}. By leveraging LoRaWAN's long-range capabilities, users can manage devices not only within their homes but also across expansive outdoor areas, such as gardens or garages, without sacrificing connectivity. This facilitates applications like remote irrigation control in gardens, where soil moisture sensors can provide feedback to automated watering systems, ensuring optimal plant health while conserving water.

Moreover, LoRaWAN's scalability allows homeowners to customize their systems based on specific needs. As technology evolves, users can integrate new sensors and devices without overhauling their existing infrastructure. For instance, adding air quality sensors can enable homeowners to monitor indoor pollutants, triggering ventilation systems when harmful levels are detected. This adaptability enhances user experience and promotes a healthier living environment.

Further research and development in this area will continue to enhance the capabilities of smart homes, providing users with advanced solutions tailored to their needs~\cite{conf/sensys/KumariG023,conf/networking/PandeyKGRR24,gupta2024utilizingtransferlearningpretrained,gupta2024lessonslearnedsmartcampus}. Innovations may include advanced machine learning algorithms that analyze sensor data to predict user behavior, further personalizing smart home interactions. For instance, predictive models can learn when users typically arrive home, adjusting lighting and temperature settings in advance to create a welcoming atmosphere.

Additionally, enhanced security features, such as facial recognition cameras and motion sensors, can work in tandem with home automation systems to provide robust monitoring solutions, alerting homeowners to potential threats in real-time. This holistic approach to home automation not only improves safety but also integrates seamlessly into users' daily lives, making technology a natural extension of their routines.

In summary, the integration of LoRaWAN with various sensors in smart homes represents a significant advancement in home automation technology. This framework not only meets the immediate needs of users but also paves the way for future innovations that will continue to shape the smart living experience.

\section{Conclusion and discussion}
The integration of LoRaWAN technology with active and passive sensors represents a significant advancement in the development of smart home systems, addressing the diverse needs of modern living while promoting sustainability and efficiency. This research has demonstrated that the synergy between these technologies not only enhances convenience and security but also fosters innovation across various applications.

Active sensors, such as motion detectors, cameras, and ultrasonic sensors, play a pivotal role in ensuring security and automation within smart homes. By actively emitting signals to detect environmental changes, these sensors provide real-time data that can trigger immediate responses—such as alerting homeowners to intruders or automating lighting systems based on occupancy. Their ability to interact with the environment allows for precise monitoring and control, which is essential in today's fast-paced lifestyle.

In contrast, passive sensors, including temperature, humidity, and light sensors, contribute to the efficiency and comfort of smart homes by passively collecting data from their surroundings. These sensors require minimal power and can operate for extended periods, making them ideal for applications where continuous monitoring is essential. By integrating passive sensors, homeowners can optimize resource usage, enhance indoor comfort, and contribute to energy conservation efforts.

LoRaWAN’s long-range capabilities and low power consumption enable these sensors to function effectively even in challenging deployment scenarios, such as remote agricultural settings or large residential properties. The ability to transmit data over long distances while maintaining a low energy profile allows for the creation of extensive sensor networks that can cover vast areas without frequent maintenance or battery replacements. This feature is particularly crucial for applications in smart agriculture and environmental monitoring, where sensor deployment may be limited by accessibility to power sources.

However, the integration of these technologies also presents challenges, such as potential interference and the need for reliable connectivity. To mitigate these issues, the design and implementation of robust communication protocols are essential. This paper has underscored the importance of selecting appropriate sensor types and communication strategies to create a resilient smart home system capable of adapting to evolving user needs and environmental conditions.

Looking ahead, further research and development in this domain will likely lead to even more sophisticated solutions tailored to individual preferences. As smart home technology continues to advance, the integration of artificial intelligence and machine learning algorithms with sensor data can enhance the responsiveness and adaptability of smart systems, enabling predictive maintenance, automated decision-making, and improved user experiences.

In conclusion, the fusion of LoRaWAN technology with active and passive sensors is paving the way for a new era of smart living that prioritizes convenience, security, and sustainability. By embracing these advancements, homeowners can enjoy a more connected and efficient living environment that not only meets their current needs but also anticipates future demands. This research lays the groundwork for future innovations in smart home systems, ultimately contributing to the broader vision of smart cities and sustainable living.
\bibliographystyle{IEEEtran}
\bibliography{paper.bib}

\end{document}